\documentclass[aps,prl,amsmath,amssymb,12pt,onecolumn,superscriptaddress,notitlepage,nofootinbib,]{revtex4-2}

\usepackage{graphicx}
\usepackage{dcolumn}
\usepackage{bm}
\usepackage{amsmath,amssymb}
\usepackage{braket}
\usepackage{wrapfig}
\usepackage{xcolor}
\usepackage{float}
\usepackage{multirow}
\usepackage{mathtools}
\usepackage{footnote}
\usepackage{comment}

\begin{document}
\title{Non-resonant cavity for intensity buildup of multiple lasers}
\author{Yi Zeng, and Nicholas R. Hutzler}
\affiliation{California Institute of Technology, Department of Physics, 1200 East California Boulevard, Pasadena, CA 91125 USA}

\begin{abstract}
A non-resonant cavity to build up laser intensity is modeled, developed and tested. It can be used for overlapping multiple lasers of different wavelengths, increasing their intensities by over an order of magnitude while maintaining good uniformity. It is simple to set up, has flexible optical characteristics, and is robust against perturbations. The intensity buildup requires no resonances, and the wavelength dependence of the performance is limited only by the mirror coatings. The cavity can be used in applications requiring a spatially-constrained intensity buildup, for example in atomic and molecular traps.
\end{abstract}

\maketitle

\section{Introduction}

In recent years, cold and controlled molecules have emerged as promising platforms for precision measurements\cite{Carr2009ColdMol,DeMille2017TableTop,Safronova2018AtomMol,Hutzler2020PolyRev}, quantum simulations\cite{DeMille2002,Barnett2006,Gorshkov2011,Bohn2017Review,Blackmore2018}, and fundamental chemistry~\cite{PrezRios2020,Bohn2017Review,Liu2022Review}.  Laser cooling and trapping, which has been a critical tool driving quantum science in atoms for decades~\cite{Metcalf1999Book}, has now been expanded to include a range~\cite{Fitch2021Review} of diatomic and even polyatomic~\cite{Kozyryev2017Sisyphus,Mitra2020CaOCH3,Vilas2022CaOHMOT} molecules.  Laser cooling of molecules remains a technical challenge, however, due in large part the difficult laser requirements; one typically needs around ten or more CW, high-power ($0.1-5$~W), narrow linewidth ($\lesssim$MHz) lasers, which must have precisely controlled and have time-varying frequency, sidebands, power, and polarizations, in order to laser cool and trap a molecule~\cite{Fitch2021Review}.  These difficulties arise due to the need to address multiple transitions in molecules, many of whose frequency separations are too large to bridge with frequency modulation or shifting.  As quantum control extends to even larger and more complicated molecules~\cite{Augenbraun2020ATM,Hutzler2020PolyRev,Mitra2022Arenes}, these requirements will become even more challenging.

One method to ease these difficulties is to use low power lasers and build up intensity with a power build-up cavity \cite{Barnes1999Cavity,King1998Cavity,Ohara2003Raman}, where resonance is used to increase the laser power circulating in the cavity. While this technique has excellent performance, it usually requires active stabilization and can only work with a small number of lasers due to the required resonant condition. Implementing this approach with $\sim10$ lasers would be very challenging, especially since laser cooling experiments typically require multiple wavelengths, sidebands, frequency changes, multiple polarizations, etc.

We are therefore interested in non-resonant methods of building up intensity. Typically, one can use a multi-pass setup bouncing the laser beams between two or more mirrors. Such a method is generally useful if the goal is to amplify power in an extended interaction region, for example with a molecule beam \cite{Kozyryev2017Sisyphus,Shuman2010}. However, the performance is limited if high intensity and uniformity are needed in a confined region, for example the few-mm cross-sectional area of a magneto-optical trap (MOT) \cite{Fitch2021Review}. In practice it is generally difficult to have intensity increases of more than a factor of a few in such a small region.

In this manuscript, we present the design and prototyping of a multipass, non-resonant and intensity-building cavity modified from the Herriott cell \cite{Herriott1965}. Using mostly off-the-shelf parts, our test setup can achieve over an order of magnitude amplification in intensity while maintaining a uniformity comparable to that of a Gaussian beam. It is also easy to set up and tune, flexible in the size of the illumination region, and robust against perturbations.

\section{The Herriott cell and its modification}

\begin{figure}[h]
\centering
\includegraphics[width=\columnwidth]{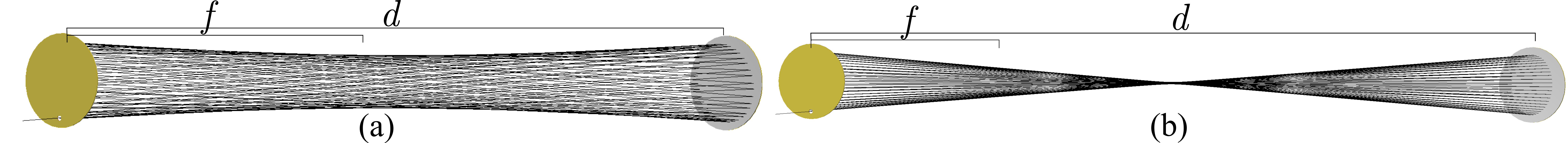}
\caption{(a) Typical Herriott cell setup in a near-confocal configuration used for multi-pass absorption spectroscopy. $d$ is the spacing between the mirrors, and $f$ is their focal length. Figure generated using LightTools. (b) Herriott cell in a near-concentric configuration.  }
\label{herifig}
\end{figure}

\begin{figure}[h]
\centering
\includegraphics[width=0.8\columnwidth]{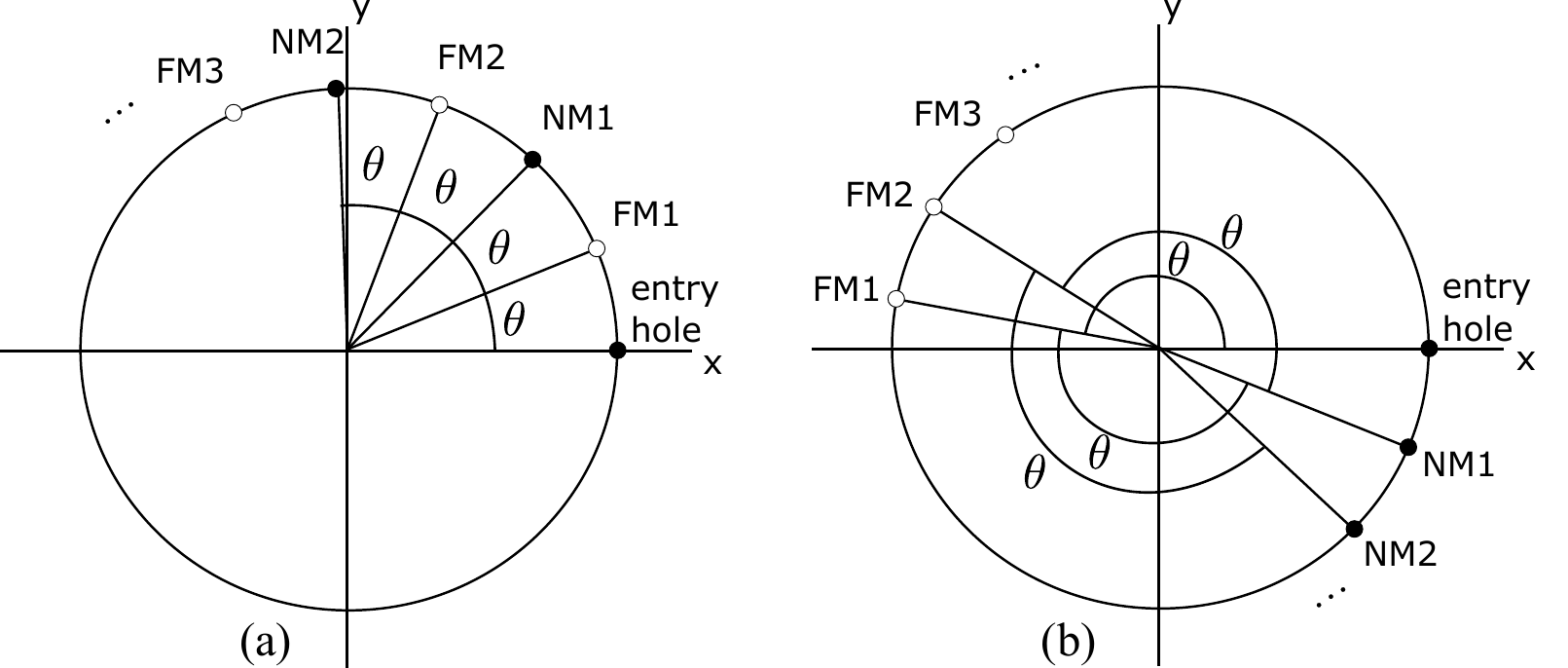}
\caption{(a) Pattern of spots traced out by the reflecting laser beam on the mirrors. The solid dots are spots on the near mirror (NM), which is the one with the entry hole, and the circles are spots on the far mirror (FM). Figure adapted from ref.~\cite{Tarsitano2007MultiHerri}. (b) Pattern when the cavity is at a near-concentric configuration, and the angle $\theta$ between consecutive spots are close to $180^{\circ}$ }
\label{dotpatfig}
\end{figure}

\begin{figure}[h]
    \centering
    \includegraphics[width=0.8\columnwidth]{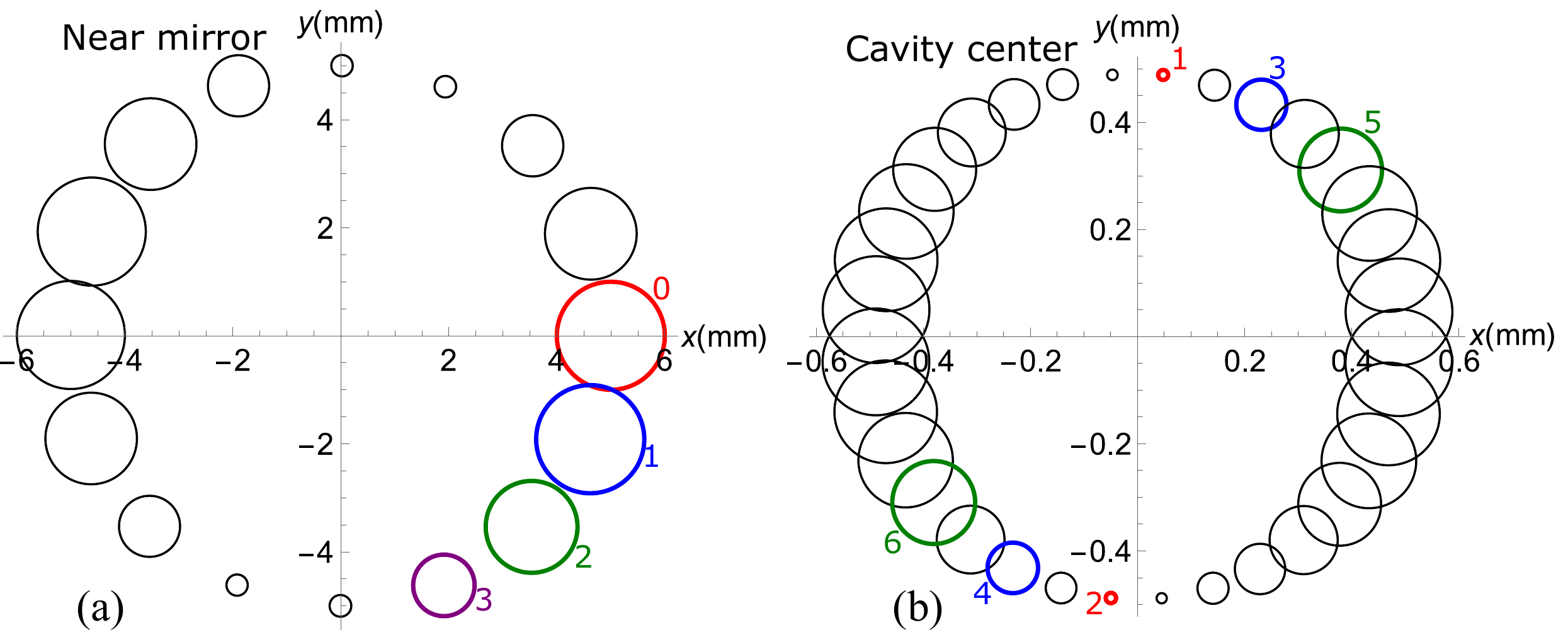}
\caption{Cross sections of laser beams in a $d=3.96f$ Herriott cell, where the spot sizes are roughly uniform, hence the ``collimated'' configuration. Here the circles indicate the size and position of the reflecting beam. The pattern is generated using a simple model based on ray transfer matrix analysis. (a) Intensity distribution on the near mirror. The entry spot 0, and first three reflecting spots are labeled. (b) Intensity distribution at the middle of the cavity; note that the size scale is 10 times smaller. The first six passes are labeled.}
\label{colisimfig}
\end{figure}

A Herriott cell is commonly used for multipass absorption spectroscopy \cite{Tarsitano2007MultiHerri,Robert2007Simple}, where the cell increases the interaction path length by factors of few tens or even hundreds, as shown in fig.~\ref{herifig}(a). The only components required are two concave mirrors, with one of them having an entry hole drilled through the face, usually near the edge. For multipass absorption, the Herriott cells are usually in a configuration where the cavity length $d$ is just slightly longer than two times the focal length $f=R/2$ of the mirrors with radius of curvature $R$, that is, nearly confocal, and the laser beam will bounce back and forth tracing out a circle of dots on the mirrors before exiting through the entrance hole. In such a configuration the laser beams form a near-cylindrical shape with little overlap, which is ideal for extending optical path lengths but not for building up intensity, which we aim to achieve by modifying the design. An easy first step would be to narrow the waist down by increasing the cavity length, as shown in fig.~\ref{herifig}. The angle between two consecutive reflecting points on the mirrors dictates the waist size, and its relationship to the cavity length is given by \cite{Herriott1965}
\begin{gather}
\cos(\theta)=1-\frac{d}{2f},
\end{gather}
where $d$ is the cavity length, $f$ is the focal length of the two mirrors, and $\theta$ is the angle between two spots on the mirrors, projected on the same plane, as shown in  fig.~\ref{dotpatfig}. As the cavity approaches the concentric configuration, $d/f$ approaches 4 and the angle approaches $180^{\circ}$ meaning that consecutive reflection points are on opposite sides so that the laser beams are always near the center at the middle of the cavity, as shown in fig.~\ref{herifig}(b) and fig.~\ref{dotpatfig}(b). The intensity distribution will look like fig.~\ref{colisimfig}, which shows calculated cross sections of laser beam sizes and positions, on the near mirror and at the middle of the cavity. The input laser beam has to be slightly focused to achieve such a ``collimated'' configuration, where the laser beam sizes on a same cross section are similar. Note that the laser beam diameter at the cavity center is about 10 times smaller than that at the cavity mirrors. 

This near-concentric configuration does not give the desired intensity buildup with uniform distribution, but we can make two further modifications to realize this goal.  First, we will move the position of the entry hole closer to the mirror center compared to the typical Herriott cell; second, we will change the divergence of the input beam.  As discussed in the following sections, these changes realize the goal of a fairly uniform intensity buildup.
\section{Modeling and prototyping}

In order to achieve a more uniform intensity distribution, we must change the input beam divergence to increase the spot sizes.  However, the stock Herriott cell mirrors used in spectroscopy usually have the entry hole very close to the mirror's edge, and we find that making the input beam more diverging results in the reflected beams leaking off the mirror edges, leading to power loss.  Thus, we would like to push the holes further in. However, we also do not want the entry hole to be too close to the center because then it will limit the number of passes the cavity can accommodate. In order to determine where the entry hole should be positioned on the near mirror and in general to better understand how the laser beam behaves when bouncing between the two concave mirrors, we implemented a simple model based on ray transfer matrix analysis \cite{saleh2019fundamentals}.

From Eq.~(1), we know where the laser beam landed on both the near and far mirrors for each pass, so the beam positions between the two mirrors can be calculated using simple geometry. Ray transfer matrix analysis is used for tracking the beam diameter of the Gaussian laser beam, which will change due to both free-space propagation and reflection from the curved mirror surfaces. The input laser beam diameter is represented by the input ray position and the focusing of the input laser is represented by the input ray angle. 

\begin{figure}[h]
         \centering
         \includegraphics[width=\columnwidth]{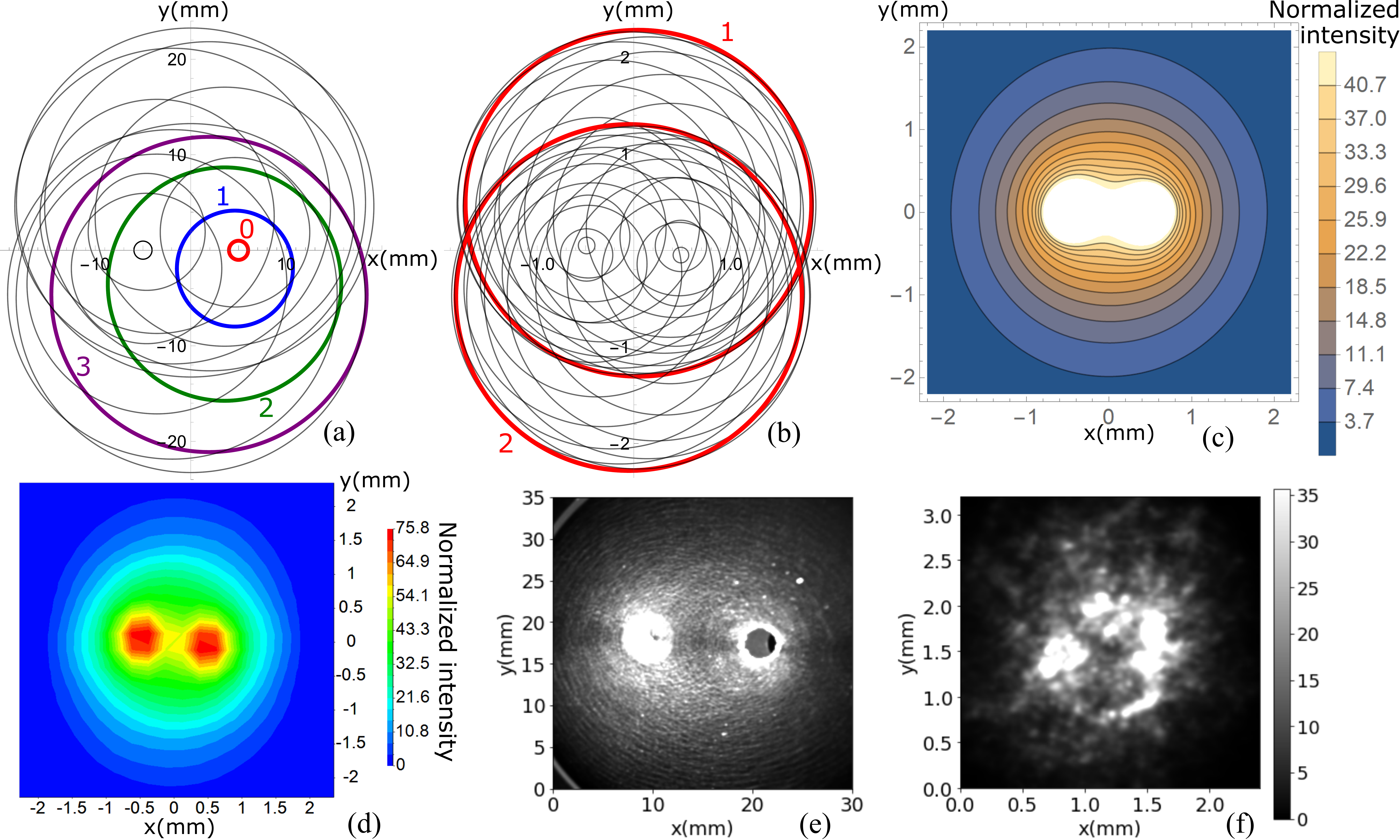}
\caption{Example of a diverging configuration. Calculated cross section patterns of laser beam sizes and positions, contour plot of intensity distribution, and photos of the same configuration in a prototype setup. (a) Pattern on the near mirror. The entry spot 0, and first three spots are labeled. (b) Pattern at the center of the cavity (size scale is 10 times smaller). The first two passes are labeled. (c) Calculated contour plot at cavity center, intensity normalized against input Gaussian beam. (d) Simulated contour plot generated from LightTools, normalized against uniform input beam. (e) Photo of the near mirror, where the bright circle on the right side is the entry hole. (f) Photo of scattered light on an AR coated window placed at cavity center, with intensity normalized against single pass. }
\label{unifullfig}
\end{figure}

Combining both the location and size information, we now have a full understanding of the laser beam inside the cavity. We can make figures like fig.~\ref{colisimfig}(a) for any cross section along the length of the cavity, and combining with a Gaussian power distribution, we can make contour plots for intensity like fig.~\ref{unifullfig}(c). We were able to generate similar plots, such as fig.~\ref{unifullfig}(d), from ray tracing simulations performed with LightTools 9.0 by Synopsys. Discrepancies between the calculated model and the simulation come from the fact that the simulation uses a input light beam of uniform intensity instead of a Gaussian beam, but it simulates light rays until they exit through either the entry hole or the mirror's edge, while the calculation only use the first orbit, 32 passes, and does not consider any leakage.

\begin{figure}[h]
\centering
\includegraphics[width=0.7\columnwidth,scale=1]{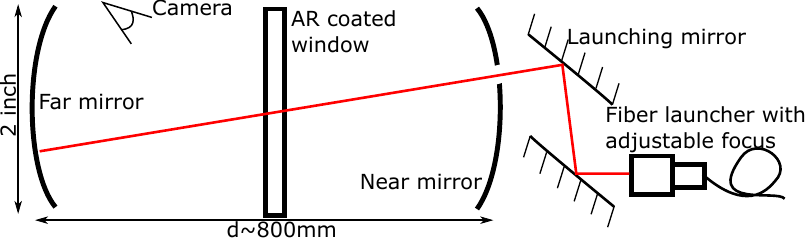}
\caption{Test setup for measuring the performance of the prototype, not to scale.}
\label{schemefig}
\end{figure}

Based on the information learned from the model, we built a prototype for testing using mostly off-the-shelf 2-inch optics. A focal length of 200 mm was chosen so that the cavity length of 792 mm, slightly less than $4f=800$~mm, would fit outside of a vacuum chamber containing atomic and molecular beams for testing. The near mirror is a custom part with a 4 mm diameter hole drilled 5 mm away from the center of the mirror\footnote{We used a Thorlabs CM508-200-E02 with the hole added by Advanced Optics, Pewaukee, WI}. We have the two cavity mirrors, and the two launching mirrors on kinematic mounts while the laser light comes out of a fiber launcher with adjustable focus close to the entry hole, as shown in fig.~\ref{schemefig}. We also insert an anti-reflection(AR)-coated window into the cavity to image the beams while minimally perturbing the paths and intensities.

The entry hole choice was not based on rigorous optimization, but it strikes a good compromise between performance, flexibility and ease of setting up and aligning. The position and size of the entry hole limit how close the first spot back on the near mirror can be to the entry spot. For us, the largest $\theta$ possible is about $169^{\circ}$, which allows for about 32 passes per orbit, so even assuming entire beam exits through the entry hole after an orbit, there is a 32-fold increase of total power. Of course, our goal is not only higher power, but also higher intensity with relatively uniform distribution. In that case, a configuration like the one shown in fig.~\ref{unifullfig} will be used. The input beam is diverged slightly, such that the first spot back on the near mirror is much larger than the entry hole. While some of the light will exit through the hole, it is small enough to be justifiable by the gain in uniformity, as shown in later sections.

From the prototyping experience, we learned that the procedure for setting up this Herriott cell multi-pass is straightforward. Put the two mirrors on kinematic mounts spaced one cavity-length apart, which is typically just short of four times the focal length, about 792 mm for our prototype. Send in the collimated laser beam, sized just under the entry hole, and land it roughly 169 degrees away from the entry hole projection at the far mirror. Adjust the far mirror orientation so that the first bounce-back to the near mirror almost touches the entry hole, like in fig.~\ref{colisimfig}(a). Finally, adjust the near mirror so that the circular patterns of dots appear. In the process, some adjustment of the focusing of the laser beam might be needed to achieve the ``collimated'' configuration, just so that we have a clear pattern of dots as indicators. Also, if the cavity length is longer than four times the focal length, the final step of forming the circular pattern will fail, since the spots will always move towards the mirror's edge and never curve back. The solution is simply to push the mirrors slightly closer, so that they can curve back to form a circle. The process is fairly robust, without needing any precise placement.

\begin{figure}[h]
         \centering
         \includegraphics[width=\columnwidth]{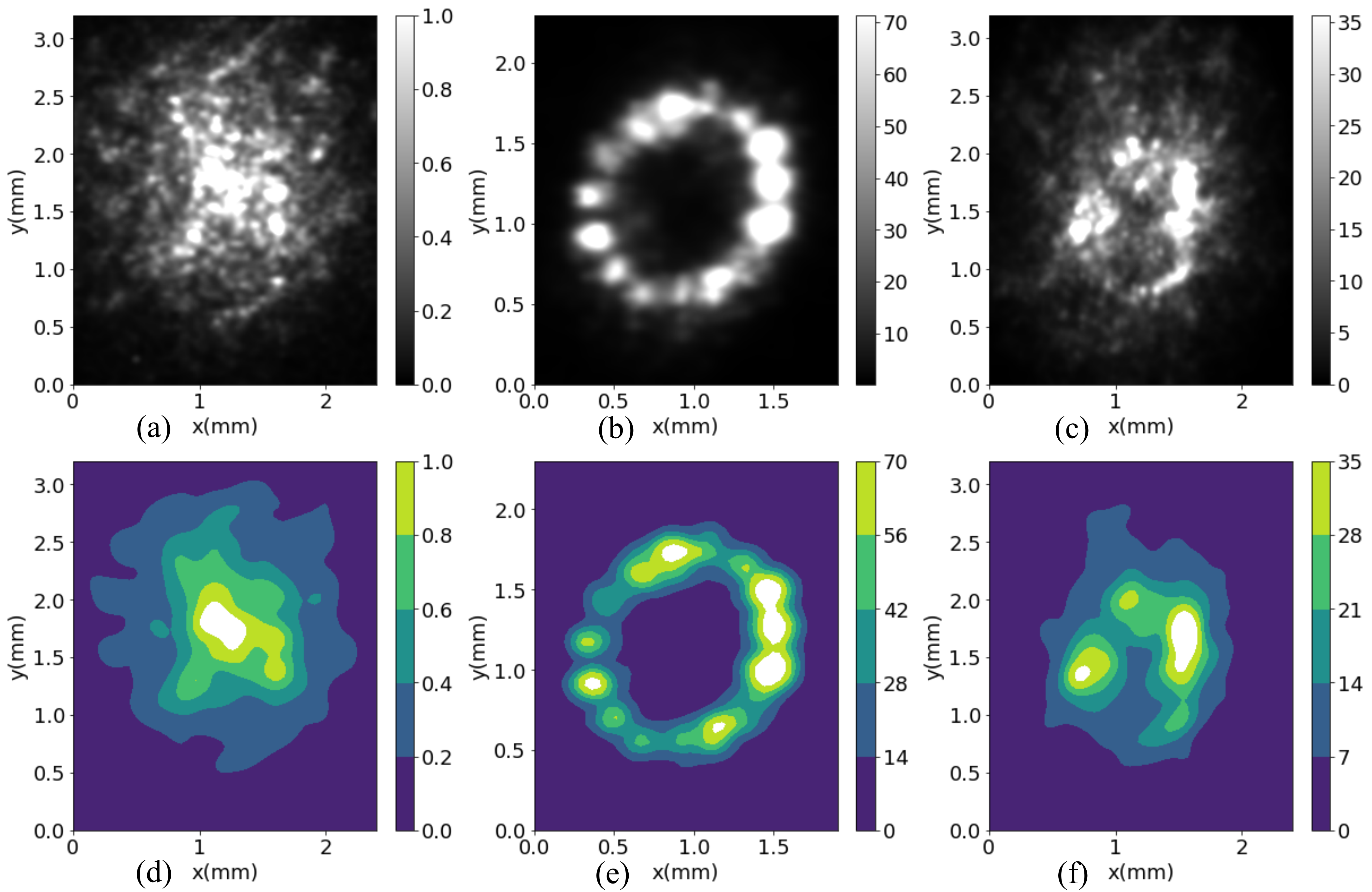}
\caption{Measurement of intensity amplification and distribution. Photos of the scattered light on the AR coated window at the center of the cavity were taken for different configurations using a CMOS camera. (a) is from a single pass of the laser beam and (d) is the contour plot of the same photo. (b) and (e) are photos for the ``collimated'' configuration. (c) and (f) are for the uniform configuration as shown in fig.~\ref{unifullfig}. All intensities are normalized against (a).}
\label{camcontfig}
\end{figure}

After achieving the ``collimated'' configuration, diverge the input beam slightly to get a configuration like that in fig.~\ref{unifullfig}. The fact that the cross section at the middle of the cavity has the same shape as that on the mirrors is a very beneficial trait for setting up and optimizing because we can reliably infer the spot characteristics inside the cavity easily just by looking at the mirrors. For optimization, the main concerns are the total power and its distribution. By using a camera or simply by looking at the mirrors, we can optimize by tuning the parameters like cavity length, entry beam focus, or mirror orientations. The total spot size at the illumination region can be changed by changing the longitudinal distance of that region from the middle of the cavity. Further discussion is in the later flexibility discussion.

\section{Performance discussion}

To quantify the performance of the setup, we inserted an AR coated window in the middle of the cavity, and measured the light scatter using a camera, as shown in fig.~\ref{schemefig}. We normalized the signal by comparing it to the scattered light from a single pass of the laser beam expanded to a similar size. The results are in good agreement with our simple model and ray tracing simulations, both for the spot shape as well as total power. The results are shown in fig.~\ref{unifullfig}(e,f) and fig.~\ref{camcontfig}. The shape of the spot patterns matches the prediction for all the configurations we tested. While the intensity results are slightly lower than simulations, it can be improved using higher-grade optics and coatings. Despite the losses, we still see that on average the intensity is amplified by a factor of about 30, with a fairly uniform distribution.

\begin{figure}[h]
         \centering
         \includegraphics[width=0.8\columnwidth]{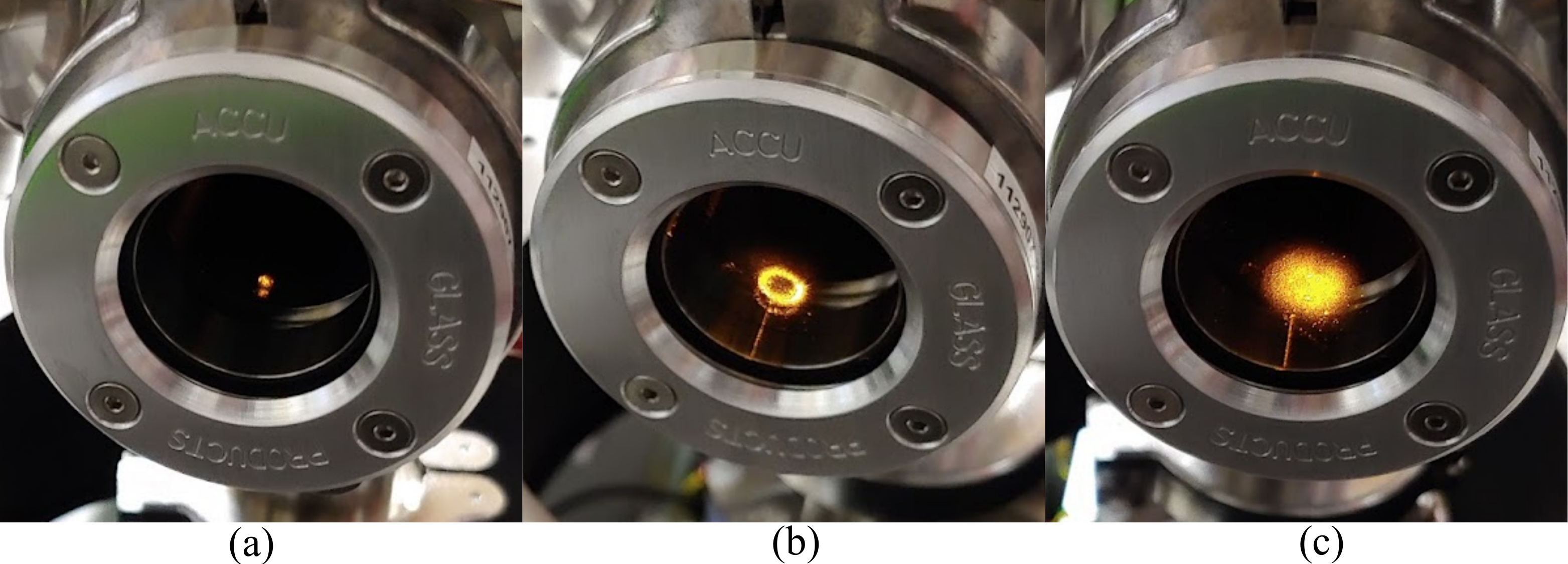}
\caption{(a) Photo of single-pass light scattered off vacuum chamber window. (b) Photo for the ``collimated'' Herriott cell setup. (c) Photo for a Herriott cell setup optimized for even intensity distribution.}
\label{vacfig}
\end{figure}

\begin{figure}[h]
\centering
\includegraphics[width=0.5\columnwidth]{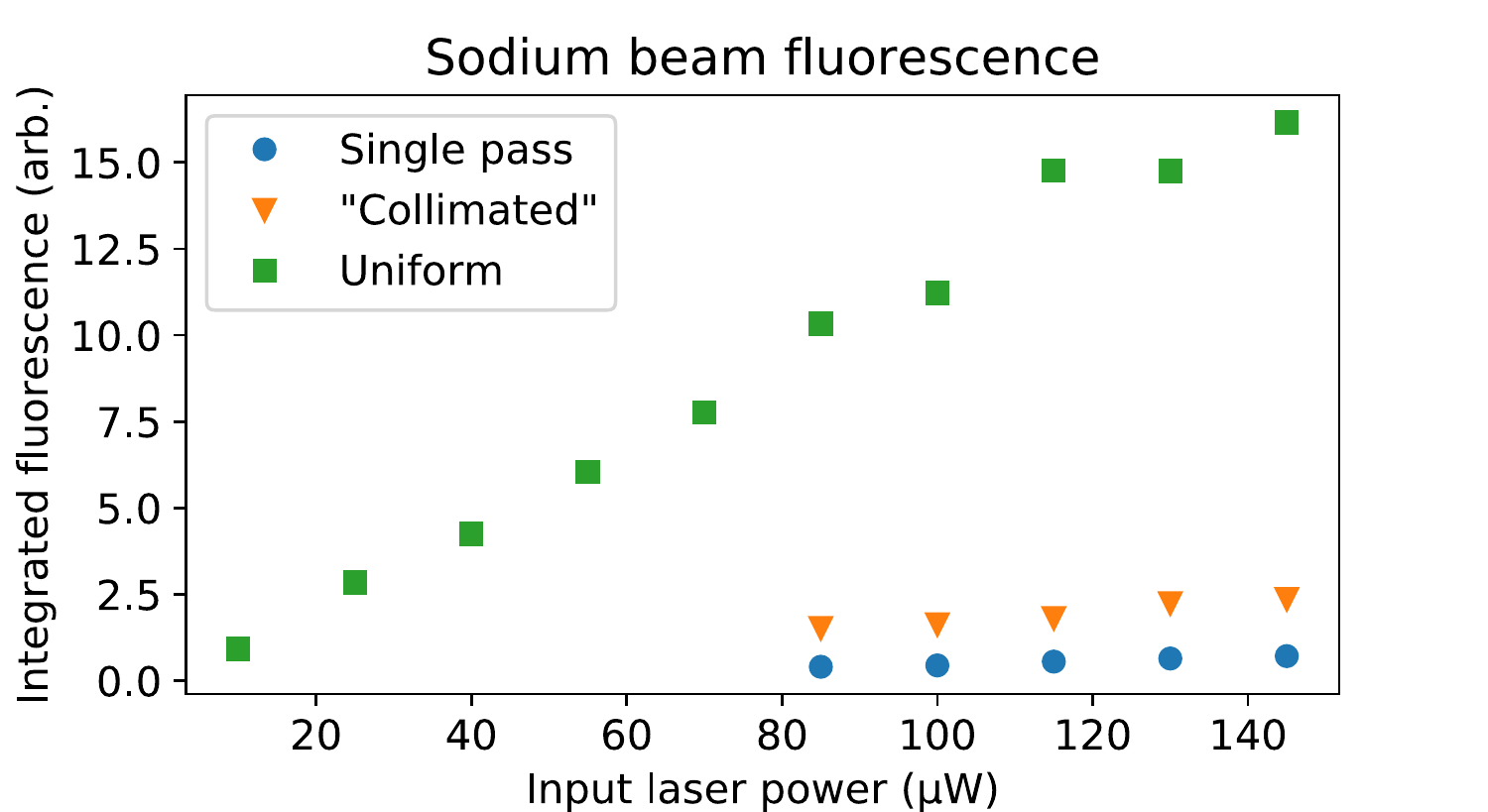}
\caption{Integrated fluorescence of a sodium beam probed on the D1 transition, comparing results from three different configurations: single pass, ``collimated'' Herriott cell, and diverging configuration like fig.~\ref{unifullfig}.}
\label{sodiumfig}
\end{figure}

Another test was to set up the cavity at our cryogenic atomic and molecular beam source \cite{Hutzler2012CBGB}, and measure the fluorescence of sodium beams using a laser on the D1 transition. Comparing the results from three configurations: 1. a single pass, as shown in fig.~\ref{vacfig}(a); 2. the Herriott cell multi-pass in the ``collimated'' configuration, as shown in fig.~\ref{vacfig}(b); 3. a diverging configuration optimized for even intensity distribution, as shown in fig.~\ref{vacfig}(c). Using the diverging Herriott cell as a power buildup cavity results in an increase in atomic fluorescence by a factor of around 25, as shown in fig.~\ref{sodiumfig}. This is less than the factor of 30 that might be expected from the intensity increase, mostly due to saturation of the atomic transition, geometry change of the fluorescing atom cloud, and power loss on the vacuum chamber windows. Otherwise, it shows that we can easily get more than one order of magnitude improvement of fluorescence signal using the prototype cavity.

\section{Flexibility and robustness}

\begin{figure}[h]
         \centering
         \includegraphics[width=0.7\columnwidth]{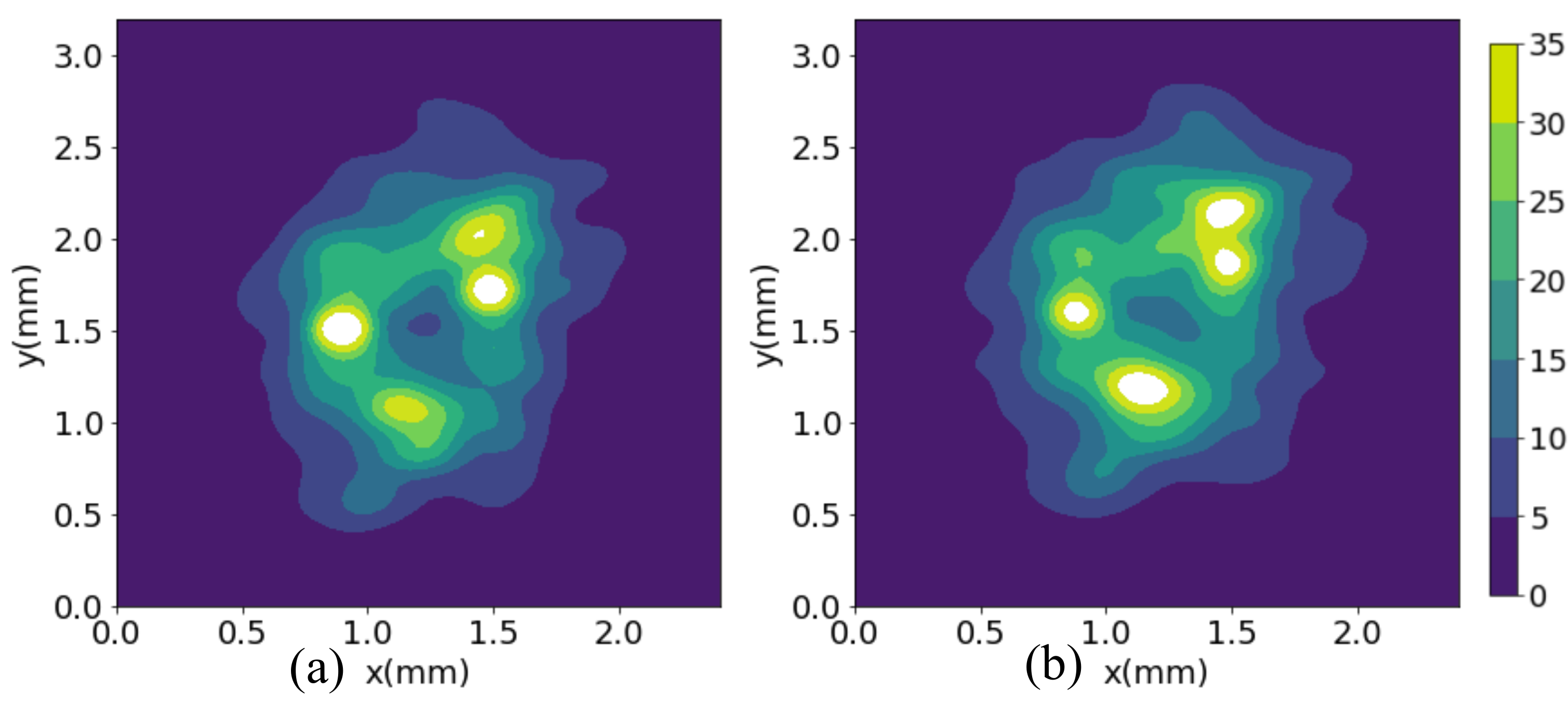}
\caption{(a) Photos of the AR coated window placed at the center of the cavity. Camera is shooting at an angle of about $45^{\circ}$, such that the scattered light from two sides of the window are sufficiently separated. (b) Same as (a) but the laser used is changed from 650 nm to 577 nm.}
\label{colorfig}
\end{figure}

One of our main goals with the cavity is the ability to build up intensity for multiple lasers of different wavelengths at the same place. Fortunately, since reflective elements naturally have no chromatic aberration, we do not have to worry about the cavity itself. Inevitably, however, the full setup will require some chromatic element in the beam path. To test if chromatic aberration or other wavelength dependent phenomena are of concern, we tried different lasers launched using the same fiber and aspheric lens. As shown in fig.~\ref{colorfig}, the intensity distributions are very similar between scattered light from a 650 nm laser and a 577 nm laser. This is expected, since the setup is very robust against small perturbations as we will discuss, especially if the perturbation is on the input laser, and not on the cavity itself.

Another very important aspect of our goal is to have an illumination region of adjustable size. The simplest way to change the size of the illuminated spot at the interaction region is by moving the entire cavity lengthwise. As mentioned in the previous section, the cross-sectional shape at the center of the cavity is the same as the pattern on the mirrors, just rotated and of different sizes. It is the case for everywhere in between as well, with the size scaled with the position. We can estimate the size of the illuminated region by noticing that the shape of the ray density roughly looks like a cone (see fig.~\ref{herifig}(b)) and should therefore be linear in displacement along the cavity:
\begin{gather}
D(z)\approx  4x_0 \left( a + 2 z \frac{1-a}{d} \right),
\end{gather}
where $z$ is the longitudinal distance of the illumination region from the middle of the cavity, $x_0$ is the offset of the entry hole from the center of the mirror, and $a=\pi-\cos ^{ - 1}(1-d/2f)$ is the angle between two consecutive laser spots    on the same mirror. The estimated diameter $D$ of the sphere being covered is twice the diameter of the circle traced out by the beam locations, which is the minimum for diverging configurations. Hence, the estimation is a conservative one. Further increasing the divergence of the input beams will increase the spot size and make the intensity distribution more uniform, though depending on the cavity geometry it might lead to more power loss over the edge of the mirrors.

\begin{figure}[h]
         \centering
         \includegraphics[width=0.5\columnwidth]{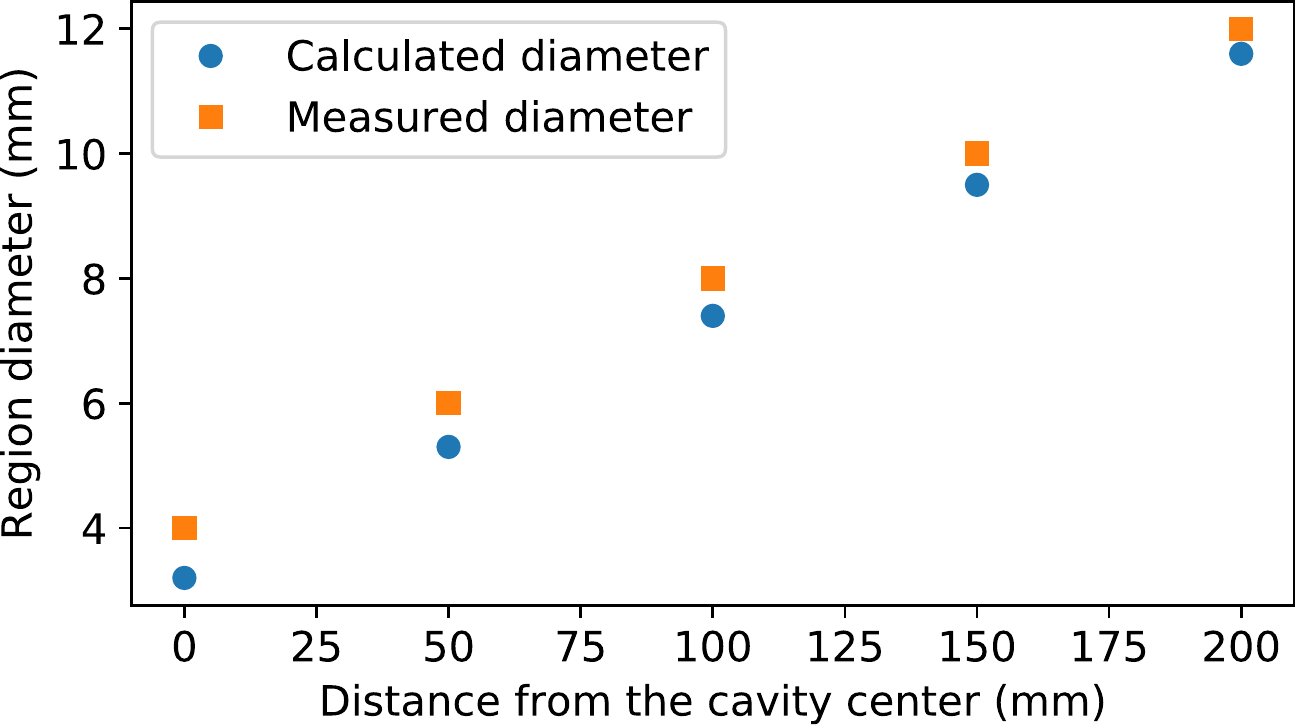}
\caption{Plot showing how the illumination region size changes with the longitudinal distance from the middle of the cavity. The measured size is characterized by the diameter of the cross section where the laser light intensity is higher than the single-pass intensity. The calculated diameter is from Eq.~(2)}
\label{sizevdistfig}
\end{figure}

To test the formula, we measured the size of the illuminated region at different locations along the cavity length. Same measurement setup was used, except the camera is imaging the window with a $~25^\circ$ angle, while moving along with the window such that the distance and angle to the window is fixed. Fig.~\ref{sizevdistfig} shows good agreement between the measured sizes and the ones calculated from Eq.~(2). 

Finally, we measured the robustness of the cavity against misalignment. In general, the cavity is very robust against small misalignment that might be caused by thermal drifts, vibrations, or even accidental bumps. For the configuration with more uniform intensity distribution, it is even less prone to misalignment, because in such a configuration the laser beams are expanded to be significantly larger than the entry hole when they arrive at the mirrors, as shown in fig.~\ref{unifullfig}, such that when they leak out of the mirror due to misalignment, there is little loss of power.

\begin{figure}[h]
         \centering
         \includegraphics[width=0.5\columnwidth]{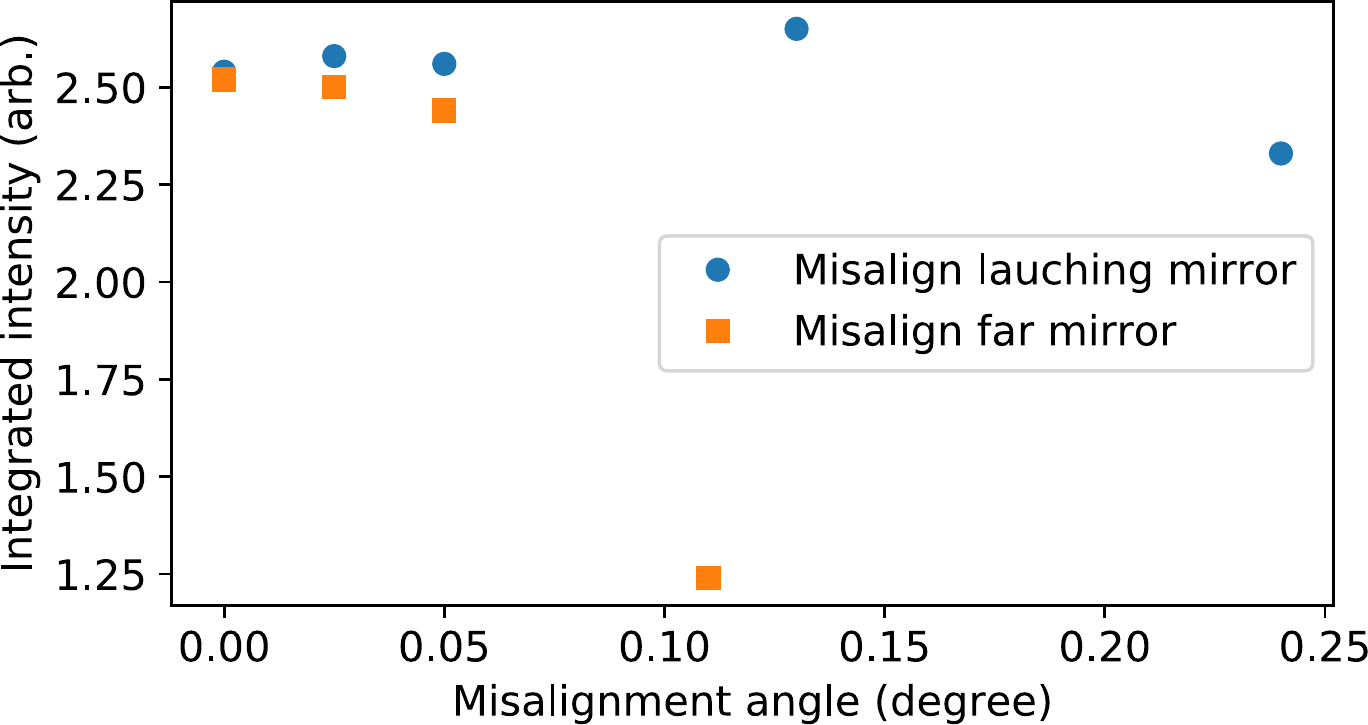}
\caption{Plot showing how misalignment in the launching mirror and far mirror affect the total power inside the illumination region. Significant power loss starts to occur when the launching mirror is misaligned by $0.24^\circ$, and when the far mirror is misaligned by $0.05^\circ$. Both are larger than typical drifts seen in lab for common optics elements.  }
\label{misalignfig}
\end{figure}

To quantify the sensitivity to misalignment, we measured how the scattered light on the AR coated window changes depending on misalignment angles of the mirrors. As expected, turning the cavity mirrors has a much more significant effect than the launching mirror as it changes the cavity condition and deviations accumulate as the laser beam bounces between them. Fig.~\ref{misalignfig} shows how the total power changes with the misalignment angles. 

The launching mirror misalignment angles reflect how the input beam is misaligned. When it is misaligned by more than $~0.24^\circ$ the total power within the cavity starts to decrease. The far mirror misalignment has the same effect as the near mirror, both misaligning the cavity. Beyond a limit of $~0.05^\circ$, the power loss starts to become significant. Clearly the cavity itself is more susceptible to misalignment than the input laser. Still, both misalignment limits are larger than the typical drifts commonly encountered in the lab; for example, the $~0.24^\circ$ misalignment required a half turn of the steering knob on the launching mirror.

\section{Conclusion}
We designed and tested a non-resonant cavity for building up laser intensity by over an order of magnitude in a confined region. It is capable of accommodating multiple lasers of different wavelengths and polarizations at the same time and at the same location, and is flexible in the size and shape of the illumination region. Furthermore, it is easy to set up and tune, and is robust against perturbations without active stabilization. These properties make it a very useful and versatile tool for laser intensity buildup, or even as an alternative to common multi-pass setup, especially when multiple laser wavelengths are required.

\subsection{Funding}
This work was supported by the Heising-Simons Foundation (2022-3361), the Gordon and Betty Moore Foundation (GBMF7947), and the Alfred P. Sloan Foundation (G-2019-12502).
\subsection{Acknowledgements}
We thank Tim Steimle, our spectroscopy collaborator, for suggesting that we look into the Herriott cell. We also thank our lab members Arian Jadbabaie and Phelan Yu for  fruitful discussions.
\subsection{Disclosures}
A draft of this manuscript was used in a provisional patent application.

\bibliography{references}

\end{document}